\documentclass[10pt]{IEEEtran}

 \usepackage{graphicx}
 \graphicspath{{eps/}} \DeclareGraphicsExtensions{.eps}

\usepackage{url}        

\usepackage{amsmath}    
\usepackage[all]{xy}
\usepackage{calc}
\usepackage{epic,eepic}
\usepackage{amssymb}
\usepackage{xcolor}

\newtheorem{theo}{Theorem}
\newtheorem{prop}[theo]{Proposition}

\newtheorem{coro}[theo]{Corollary}

\newtheorem{exam}[theo]{Example}

\renewcommand{\H}{{\cal H}}
\newcommand{\N}{{\cal H}}
\newcommand{\M}{{\cal H}}

\newcommand{\pr}{\mbox{\rm Pr}}

\newcommand{\sgn}{{\mbox{\rm sgn}}}
\newcommand{\tmp}{{\mbox{\rm\tiny tmp}}}

\renewcommand{\min}{\mathop{\mbox{\rm min}}}

\newcommand{\topspace}[1][4.25mm]{\vbox{\hbox{\vspace{#1}}}}

\newenvironment{mylist}[1][$\bullet$]{\begin{list}{#1}{\itemindent .5\parindent \leftmargin 0mm }}{\end{list}}
\newcommand{\myitem}{\item}

\begin{document}
%
\title{Self-Corrected Min-Sum decoding of LDPC codes}
%
%
\author{Valentin~Savin, CEA-LETI, MINATEC, Grenoble, France, valentin.savin@cea.fr
\thanks{Part of this work was performed in the frame of the %
ICTNEWCOM++ project which is a partly EU funded Network of %
Excellence of the 7th European Framework Program.}
}
\maketitle

\begin{abstract}
In this paper we propose a very simple but powerful self-correction
method for the Min-Sum decoding of LPDC codes. Unlike other
correction methods known in the literature, our method does not try
to correct the check node processing approximation, but it modifies
the variable node processing by erasing unreliable messages.
However, this positively affects check node messages, which become
symmetric Gaussian distributed, and we show that this is sufficient
to ensure quasi-optimal decoding performance. Monte-Carlo
simulations show that the proposed Self-Corrected Min-Sum decoding
performs very close to the Sum-Product decoding, while preserving
the main features of the Min-Sum decoding, that is low complexity
and independence with respect to noise variance estimation errors.
\end{abstract}


%
\IEEEpeerreviewmaketitle

\section{Introduction}
%
%
%
%
Low density parity check (LDPC) codes can be iteratively decoded using message-passing type algorithms that may be classified as optimal,
sub-optimal or quasi-optimal. The optimal iterative decoding is performed by the Sum-Product algorithm \cite{Wiberg} at the price of an
increased complexity, computation instability, and dependence on thermal noise estimation errors. The Min-Sum algorithm \cite{Wiberg}
performs a sub-optimal iterative decoding, less complex than the Sum-Product decoding, and independent of thermal noise estimation errors.
The sub-optimality of the Min-Sum decoding comes from the overestimation of check-node messages, which leads to performance loss with
respect to the Sum-Product decoding. However, as we will highlight in this paper, it is not the overestimation by itself, which causes the
Min-Sum sub-optimality, but rather the loss of the symmetric Gaussian distribution of check-node messages.

Several correction methods were proposed in the literature in order to recover the performance loss of the Min-Sum decoding with respect to the
Sum-Product decoding. We call such algorithms quasi-optimal. Ideally, a quasi-optimal algorithm should perform very close to the
Sum-Product decoding, while preserving the main features of the Min-Sum decoding, that is low complexity and independence with respect to
noise variance estimation errors. The most popular correction methods are the Normalized Min-Sum and the Offset Min-Sum algorithms
\cite{Chen-Foss}. These algorithms share the idea that the performance loss can be recovered by reducing the amplitudes of
check-node messages, which is desirable as these messages are actually overestimated within the Min-Sum decoding.

In \cite{sav:nr} we showed that the performance loss of the Min-Sum decoding can be recovered by ``cleaning'' the computation trees
associated with the decoding process.
While this was the first example of a self-correction method of the Min-Sum decoding, the drawback is that it requires  a different
processing order of nodes at each iteration, which may be hardly implementable in hardware.

The starting point of this paper was to figure out how to benefit from the idea of ``cleaning'' computation trees without requiring any
particulary processing order of nodes. The proposed {\em Self-Corrected Min-Sum} decoding detects unreliable information simply by the sign
fluctuation of variable node messages. Precisely, any variable node message changing its sign between two consecutive iterations is {\em
erased}, meaning that any such a fluctuating message is set to zero. The Self-Corrected Min-Sum decoding will be analyzed from two
different but complementary points of view. They can be summarized as follows:

\begin{mylist}
\item Erasing fluctuating messages may be seen as cleaning computation trees associated with the decoding process of some
unreliable information. We will show that the Self-Corrected Min-Sum decoding behaves as the Min-Sum decoding applied on a sub-tree of
unerased messages.
\item We also show that the check node messages exchanged within the Self-Corrected Min-Sum decoding have a symmetric Gaussian
distribution, and that this is sufficient to ensure quasi-optimal
decoding performance.
\end{mylist}

The paper is organized as follows. In the next section we introduce and explain the Self-Corrected Min-Sum algorithm. In section
\ref{sec:computation_tree} we study computation trees associated with the decoding process, and describe the behavior of the Self-Corrected
Min-Sum decoding on graph with cycles. In section \ref{sec:analysis} we show that check node messages have a symmetric Gaussian
distribution and we discuss the analysis of the Self-Corrected Min-Sum decoding by Gaussian approximation. Finally, section
\ref{sec:simulations} presents simulation results and section \ref{sec:conclusions} concludes this paper.

The following notations concern bipartite graphs and message-passing algorithms running on these graphs, and will be used throughout the
paper.

\begin{mylist}
\myitem $\H$, the Tanner graph of a LDPC code, comprising $N$ variable nodes and $M$ check nodes, 
\myitem $n \in \{ 1, 2, \dots, N \}$, a variable node of ${\cal H}$,
\myitem $m \in \{ 1, 2, \dots, M \}$, a check node of ${\cal H}$, 
\item ${\cal H}(n)$, set of neighbor check nodes of the variable node $n$,
\item ${\cal H}(m)$,\!\hfill set\!\hfill of\!\hfill neighbor\!\hfill variable\!\hfill nodes\!\hfill of\!\hfill the\!\hfill check\!\hfill node\!\hfill
$m$,
\item $\gamma_n$, a priori information (LLR) of the variable node $n$,
\item $\tilde{\gamma}_n$, a posteriori information (LLR) of the variable node $n$,
\item $\alpha_{m,n}$, the variable-to-check message from $n$ to  $m$,
\item $\beta_{m,n}$, the check-to-variable message from $m$ to $n$.
\end{mylist}

\pagebreak

\section{Self-Corrected Min-Sum decoding}

The Self-Corrected Min-Sum decoding performs the same initialization step, check node processing, and a posteriori information update as
the classical Min-Sum decoding. The variable node processing is modified as shown below:

\smallskip\noindent {\bf Initialization} \\
{\it $\bullet$ A priori information}
$$\gamma_n(a) = \ln\left({\pr( x_n = 0 \mid \mbox{channel} )} /
                               {\pr( x_n = 1 \mid \mbox{channel} )}\right)$$
{\it $\bullet$ Variable-to-check messages initialization}
$$\alpha_{m,n} = \gamma_n$$

\smallskip\noindent{\bf Iterations}\\
 {\it $\bullet$ Check node processing}
$$\beta_{m,n} = \displaystyle\left(\prod_{n'\in\N(m)\setminus\{n'\}} \!\!\!\!\!\!\!\!\!\!\sgn(\alpha_{m,n'})\right)\cdot
\min_{n'\in\N(m)\setminus\{n\}}(\mid\alpha_{m,n'}\mid)$$
 {\it $\bullet$ A posteriori information}
$$\tilde\gamma_n = \gamma_n + \displaystyle\sum_{m\in\M(n)}\beta_{m,n}$$
 {\it $\bullet$ Variable node processing}
$$\begin{array}{l}
\alpha_{m,n}^\tmp = \tilde\gamma_n - \beta_{m,n} \\
\topspace[4.75mm]\mbox{\em if }\ \sgn(\alpha_{m,n}^\tmp) = \sgn(\alpha_{m,n})\ \ \mbox{\em then }\ \alpha_{m,n} = \alpha_{m,n}^\tmp\\
\topspace[4.75mm]\mbox{\em else }\ \alpha_{m,n} = 0 
\end{array}$$

The variable node processing is explained below:
\begin{mylist}
\myitem First, we compute the new extrinsic LLR for the current iteration by $\alpha_{m,n}^\tmp = \tilde\gamma_n - \beta_{m,n}$. However,
unlike the classical Min-Sum decoding, this value is stored as an intermediary (temporary) value. We note that at this moment the message
$\alpha_{m,n}$ still contains the value of the previous iteration.

\myitem Next, we compare the signs of the temporary $\alpha_{m,n}^\tmp$ (extrinsic LLR computed at the current iteration) and the message
$\alpha_{m,n}$ that was sent by the variable node $n$ to the check node $m$ at the previous iteration. If the two signs are equal we update
the variable node message by $\alpha_{m,n} = \alpha_{m,n}^\tmp$ and send this value to the check node $m$.

\myitem On the other hand, if the two signs are different, the variable node $n$ send an erasure to the check node $m$, meaning that the
variable node message $\alpha_{m,n}$ is set to zero.
\end{mylist}

It should be noted that we consider the zero message to have both negative and positive signs. In other words, {\em whenever the old
message $\alpha_{m,n} = 0$,  we update the new variable node message by $\alpha_{m,n} = \alpha_{m,n}^\tmp$}.

\begin{figure}[t]
\noindent\includegraphics[width=\linewidth]{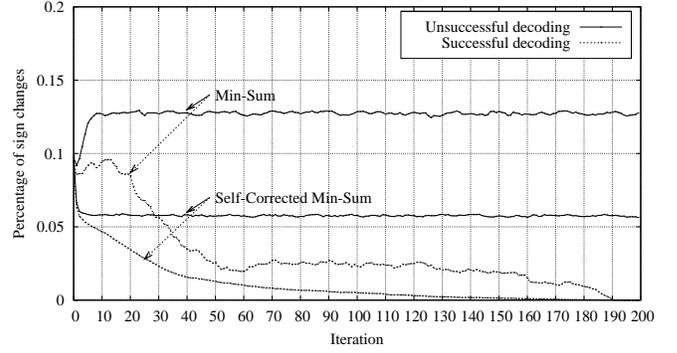}
 \caption{Percentage of sign changes per iteration: AWGN, coding rate $1/2$, $E_b/N_0 = 1$\,dB, maximum iteration number = $200$}
 \label{fig:sign_fluctuation}
\end{figure}

Fig. \ref{fig:sign_fluctuation} presents the percentage of sign changes for each decoding iteration. Obviously, this percentage tends to zero in case of successful decoding.  The dotted curves, corresponding to successful decoding, indicate that the Self-Corrected Min-Sum
decoder converges faster than the Min-Sum decoder. In case of unsuccessful decoding, it can be observed that the percentage of sign changes
tends very quickly to a positive value. At the first iteration both Min-Sum and Self-Corrected Min-Sum decodings present the same
percentage of sign changes, since in both cases variable-to-check messages are initialized from the same a priori information. Nonetheless,
they behave in a completely different way during the first iterations: the percentage of sign changes increases in case of Min-Sum
decoding, while it decreases within the Self-Corrected Min-Sum decoding. Thus, one consequence of the self-correction method is to control sign fluctuations when they occur, which help the decoder to enter into a convergence mode.

\section{Computation trees} \label{sec:computation_tree}

Computation trees proved to be very useful in understanding the behavior of message-passing decoders running on graphs with cycles
\cite{Wiberg}. Consider some information iteratively computed at some variable node $n$. By examining the updates that have occurred, one
may recursively trace back through time the computation of this information. This trace back will form a tree graph rooted at $n$ and
consisting of interconnected variable and check nodes in the same way as in the original graph, but the same variable or check nodes may
appear at several places in the tree. We denote by $\Gamma_n^{(l)}$ and ${\cal A}_{m,n}^{(l)}$ the computation trees of $\tilde\gamma_n$
and $\alpha_{m,n}$ at the $l^{\mbox{\scriptsize th}}$ iteration. Both trees are rooted at $n$, the difference is that $\Gamma_n^{(l)}$ does
contain a copy of $m$ as child node of $n$, while ${\cal A}_{m,n}^{(l)}$ does not. Moreover, we note that these computation trees do not
depend on the decoding algorithm, but only on the Tanner graph and the iteration number.

Let ${\cal T}$ be an arbitrary bipartite tree of depth\footnote{The depth of a variable node is defined recursively from root to leaf
nodes: the root node is of depth zero and the depth of any other variable node is one plus the depth of its grandparent variable node. The
depth of the tree is the maximum depth of its variable nodes.} $L$, whose root and leaf nodes are all variable nodes. If {\bf Dec} is a
message-passing decoder running on ${\cal T}$, we denote by {\bf Dec}$({\cal T})$ the a posteriori information of the root node of ${\cal
T}$ at the $L^{\mbox{\scriptsize th}}$ iteration.

\begin{exam}
Let {\bf MS} and {\bf SCMS} denote the Min-Sum and Self-Corrected Min-Sum decoders respectively. Then {\bf MS}$({\cal A}_{m,n}^{(l)})$ and
{\bf SCMS}$({\cal A}_{m,n}^{(l)})$ are equal to the variable node messages $\alpha_{m,n}$ computed by the Min-Sum and Self-Corrected
Min-Sum algorithms respectively, at the $l^{\mbox{\scriptsize th}}$ iteration.
\end{exam}

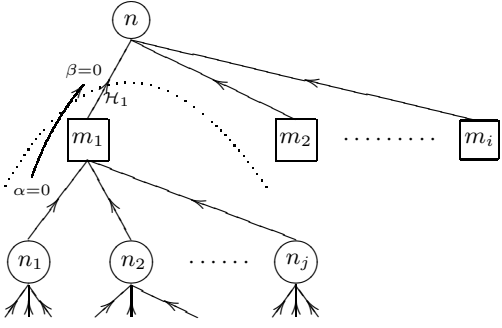
\begin{figure}[t]
  \centering
  {\small $\xymatrix@R=12pt@C=0pt{
  &  &  &  & *+=[o][F]{\,\,\,n\,\,\,} \ar@{-}'[]!D[ddl]!U|{\object@{<}}_{\beta = 0}="b"\ar@{-}'[]!D[ddrrr]!U|{\object@{<}}\ar@{-}'[]!D[ddrrrrrrr]!U|{\object@{<}} &
  &  &  & &  &  & \\
  &   &   &   &   &  &   &   &                        \\
  &  &  &  *+=[F]{\,m_1\,}\ar@{-}'[]!D[ddll]!U|{\object@{<}}_{\alpha = 0}="a" \ar@{-}'[]!D[ddr]!U|{\object@{<}} \ar@{-}'[]!D[ddrrrr]!U|{\object@{<}} &  &  &  &   *+=[F]{\,m_2\,} & &\cdots\cdots\cdots & & *+=[F]{\,m_i\,} \\
  \ar@{.}@/^3.5pc/[drrrrrrrr]+<-8mm,10mm>_(.45){{\cal H}_1}  &   &   &   &   &  &   &   & \ar@{->}@/^0.15pc/ "a"+<0pt,5pt>;"b"+<0pt,-5pt>           \\
   & *+=[o][F]{\,\,\,n_1\,\,}\ar@{-}'[]!D[dl]!UR|{\object@{<}}\ar@{-}'[]!D[d]!U|{\object@{<}}\ar@{-}'[]!D[dr]!UL|{\object@{<}} &                           &
  & *+=[o][F]{\,\,\,n_2\,\,}\ar@{-}'[]!D[dl]!UR|{\object@{<}}\ar@{-}'[]!D[d]!U|{\object@{<}}\ar@{-}'[]!D[dr]!UL|{\object@{<}} & \mbox{ }\hspace{1.8mm}\cdots\cdots&
  & *+=[o][F]{\,\,\,n_j\,\,}\ar@{-}'[]!D[dl]!UR|{\object@{<}}\ar@{-}'[]!D[d]!U|{\object@{<}}\ar@{-}'[]!D[dr]!UL|{\object@{<}} &   \\
  &  &  &
  &  &  &  &  &                                                           }$}
  \caption{Computation tree (part of)} \label{fig:tree}
\end{figure}

\begin{prop}
For any nodes $n,m$ and any iteration $l$, there exists a sub-tree ${\cal T}_{m,n}^{(l)} \subset {\cal A}_{m,n}^{(l)}$ containing $n$ such
that:
 $$\mbox{\bf SCMS}({\cal A}_{m,n}^{(l)}) = \mbox{\bf MS}({\cal T}_{m,n}^{(l)})$$
\end{prop}
The same holds for the a posteriori information tree $\Gamma_n$.

Indeed, assume that $\M(n) = \{m,m_1,\dots,m_i\}$, and let ${\cal A}_{m,n}^{(l)}$ be the computation tree represented in Fig.~\ref{fig:tree}. If none of the messages from variable nodes
$n_1,\dots, n_j$ to the parent check node $m_1$ has been erased  at iteration $l-1$, then the processing of $n_1,\dots, n_j$ and $m_1$
corresponds to classical Min-Sum updates. On the contrary, assume that at iteration $l-1$, the message sent by the variable node $n_1$ to
its parent check node $m_1$ has been set to zero. Then, at iteration $l$, the check node message sent by the check node $m_1$ to its parent
variable node $n$ is also equal to zero. Therefore, we can omit this message when processing the variable node $n$, or equivalently, the
sub-tree ${\cal H}_1$ corresponding to the check node $m_1$ can be discarded. The proposition is proved by moving down the tree and
applying the above arguments recursively.

From a more intuitive perspective, the above proposition states that {\em the Self-Corrected Min-Sum decoding behaves as the Min-Sum
decoding applied on the sub-tree of unerased messages}.

\section{Analysis of the Self-Corrected Min-Sum decoding using Gaussian approximation} \label{sec:analysis}
Throughout this section we restrict ourselves to the AWGN channel with noise variance $\sigma^2$, and we consider the BPSK modulation
mapping bit $ 0 \mapsto +1$ and bit $1 \mapsto -1$. Without loss of generality, we further assume that the all-zero codeword is sent over
the channel. At the receiver side, a message-passing algorithm is used to decode the received signal. Let $P_e^l$ be the decoding error
probability at iteration $l$. Therefore, $P_e^l$ is simply the average probability that variable node messages are non-positive:
$$P_e^l = \Pr(\alpha_{m,n}^l \leq 0)$$
where superscript $l$ is  used (and will be used from now on) to denote messages sent at the $l^{\mbox{\scriptsize th}}$ iteration. We note
that the probability $\Pr(\alpha_{m,n}^l = 0)$ is zero for the Sum-Product or the Min-Sum decoding, but this is not the case anymore for
the Self-Corrected Min-Sum decoding. The Gaussian approximation is an approach to track the error probability of a message-passing
decoding, based on Gaussian densities. The following conditions are needed for the analysis by Gaussian approximation:


\smallskip \noindent (GD) Gaussian distribution condition: messages received at every node at every iteration are independent and identically
distributed (i.i.d), with symmetric Gaussian distribution of the form:
 $$f(x) = \displaystyle\frac{1}{\sqrt{4\pi m}}e^{-\frac{(x-m)^2}{4m}}$$
 where the parameter $m$ is the mean.

 We will denote by $m_0 = 2/\sigma^2$ the mean of the a priori information, and by $m_\alpha^l$ and
 $m_\beta^l$ the means of variable and check node messages at iteration l, respectively.

\smallskip \noindent (CNP) Check node processing condition:
 $$\sgn(\beta_{m,n}^{l+1}) = \prod_{n'\in\N(m)\setminus\{n'\}} \!\!\!\!\!\!\!\!\!\!\sgn(\alpha_{m,n'}^l)$$

\smallskip \noindent (VNP) Variable node processing condition:
 $$\alpha_{m,n}^l = \gamma_n + \displaystyle\sum_{m'\in\M(n)\setminus\{m\}} \!\!\!\!\!\!\!\!\!\!\beta_{m',n}^l$$

The irregularity of the LDPC code is specified as usual, using variable and check node distribution degree polynomials:
 $\lambda(x) = \sum_{i=2}^{d_v}\lambda_ix^{i-1}$, where $d_v$ is the maximum variable node degree and $\lambda_i$ is the
fraction of edges connected to variable nodes of degree $i$, respectively $\rho(x) = \sum_{j=2}^{d_c}\rho_jx^{j-1}$, where $d_c$ is the
maximum check node degree and $\rho_j$ id the fraction of edges connected to check nodes of degree $i$

 \begin{theo} Let $\varphi$ be the function defined by:
 $$\varphi(x) = \sum_{i=2}^{d_v}\lambda_i Q\left(\sqrt{\frac{1}{\sigma_2} + (i-1)Q^{-1}\left(\frac{1-\rho(1-2x)}{2}\right)}\right)$$
 where $Q(x) =\displaystyle\frac{1}{\sqrt{2\pi}}\int_x^{+\infty} e^{-\frac{t^2}{2}} dt$.
Then $\varphi$ depends only on $\sigma$, $\lambda$, and $\rho$,  and for any message-passing decoder satisfying the conditions (DG), (CNP),
and (VNP), the following recurrence relation holds:
$$P^{l+1}_e = \varphi(P^l_e)$$
 \end{theo}

 The main consequence of this theorem is that the {\em exact} check node processing is not really important. It is only important that
 check node messages have a symmetric Gaussian distribution and their signs verify the (CNP) condition.
  The proof can be derived similarly as in \cite{lehmann_analysis_2003}, noting that only the conditions (DG), (CNP) and (VNP) are needed,
  not the Sum-Product decoder assumption by itself.

 \begin{figure*}[!t]
 \centering
 \begin{minipage}{.33\linewidth}
   \centering
   \includegraphics[width=\linewidth]{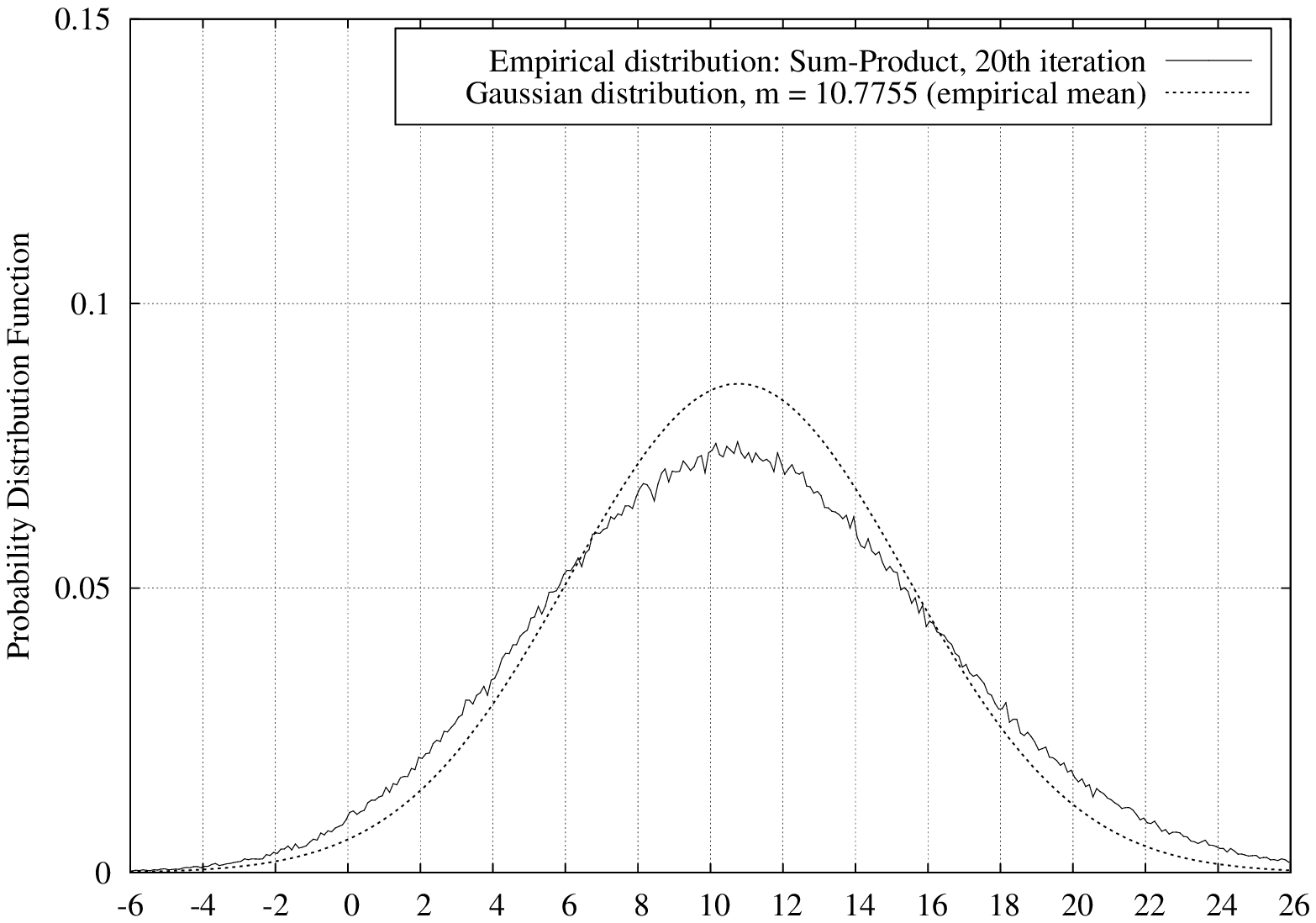}
   {\footnotesize $(a)$ Sum-Product decoding}
 \end{minipage}\hfill
 \begin{minipage}{.33\linewidth}
   \centering
   \includegraphics[width=\linewidth]{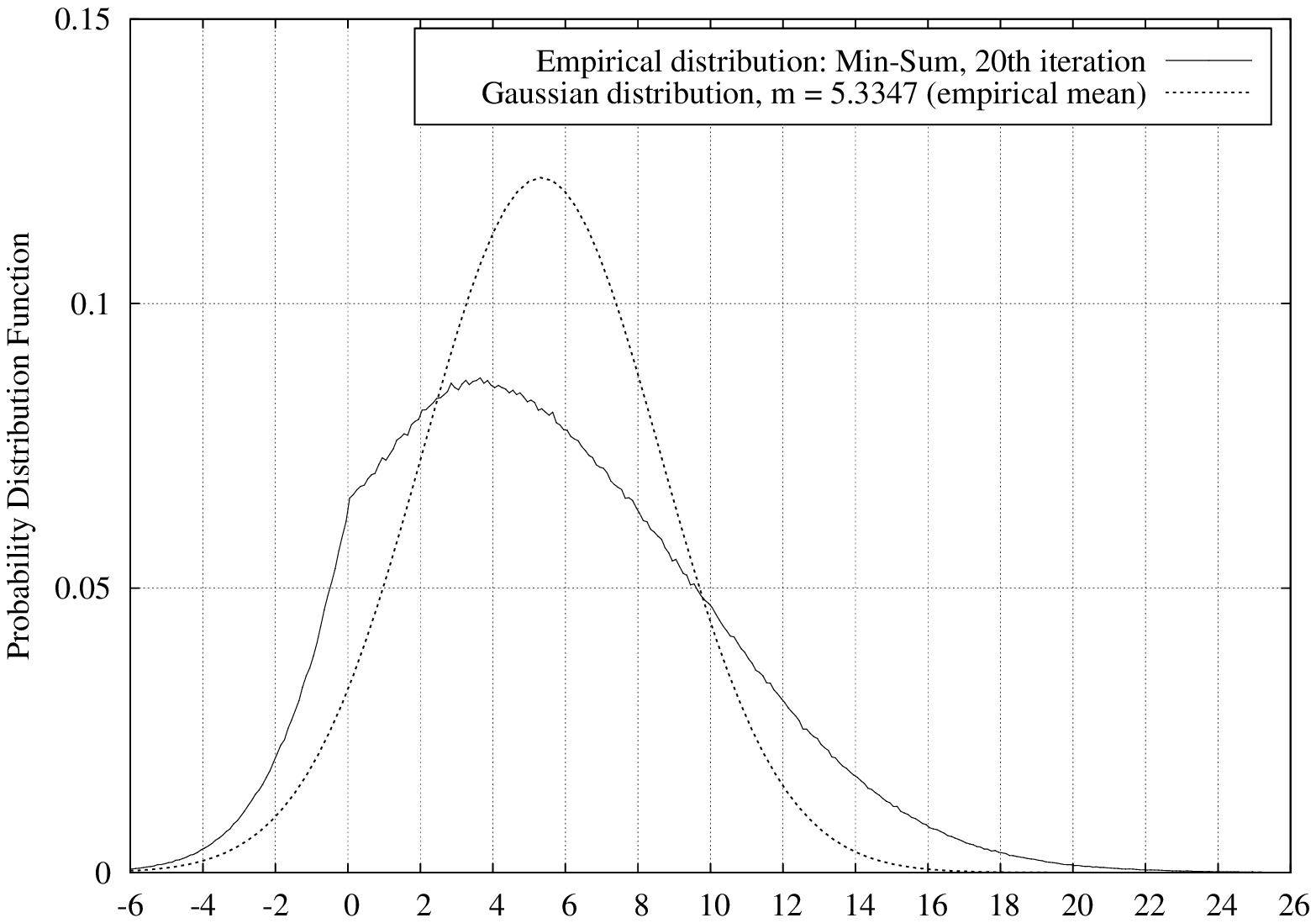}
   {\footnotesize $(b)$ Min-Sum decoding}
 \end{minipage}\hfill
 \begin{minipage}{.33\linewidth}
   \centering
   \includegraphics[width=\linewidth]{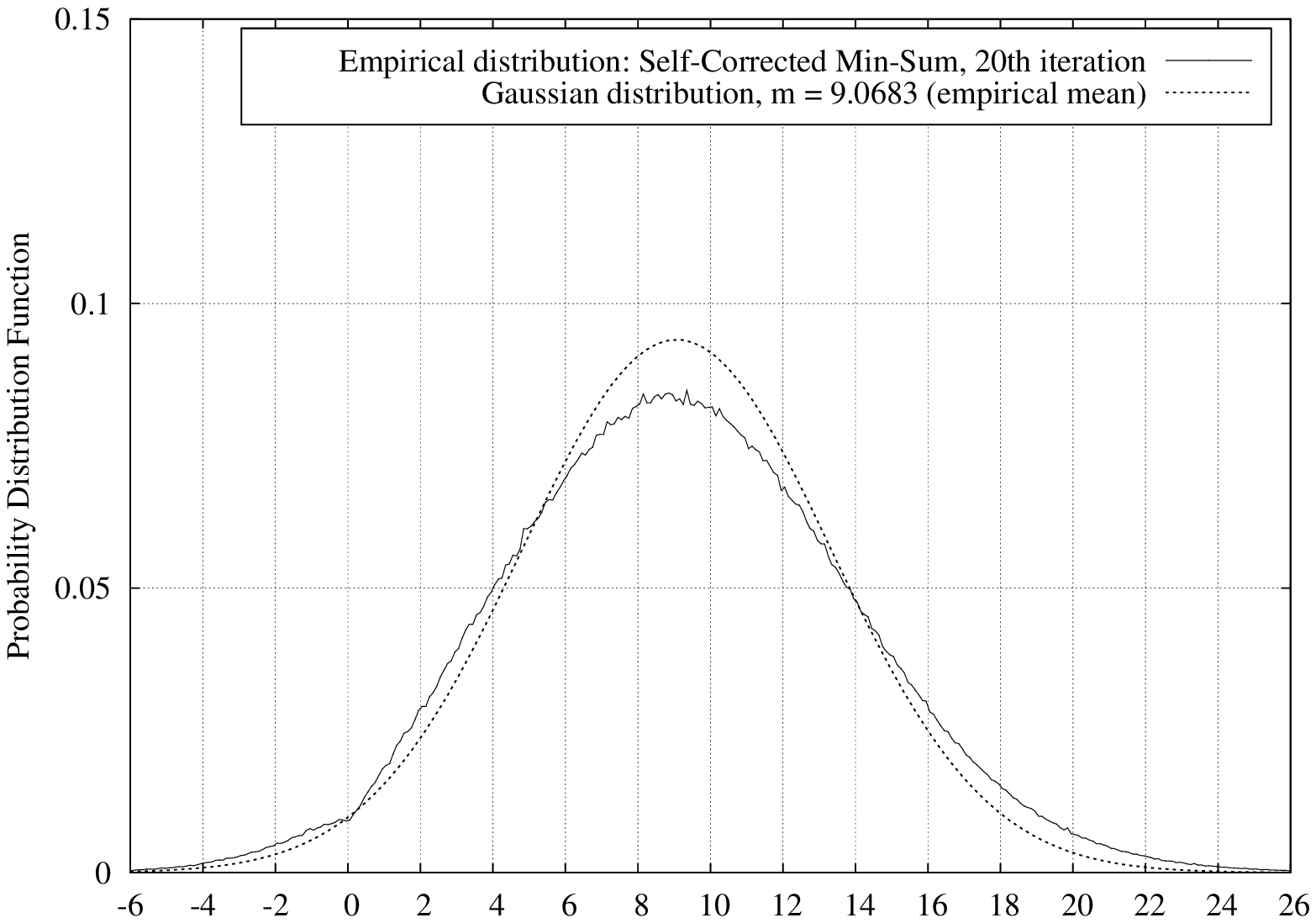}
   {\footnotesize $(c)$ Self-Corrected Min-Sum decoding}
 \end{minipage}
 \caption{Empirical probability density vs. Gaussian density after 20 iterations ($E_b/N_0 = 1.5$\,dB)}
 \label{fig:density}
 \end{figure*}

 Conditions (CNP) and (VNP) are verified by both Sum-Product and  Min-Sum decoders. The performance gap between these decoders
 is explained by the fact that the Min-Sum fails to satisfy the Gaussian distribution (GD)
 condition for check node messages. This can be seen in Fig. \ref{fig:density} $(a)$ and $(b)$ that represent the empirical probability densities
  of check node messages computed by the Sum-Product and the Min-Sum decoders. A straightforward consequence of the above theorem is that if the check
  node messages
 of the Min-Sum decoder would have a Gaussian-like distribution, then the Min-Sum and the Sum-Product decoders would exhibit the
 same performance.

 In Fig. \ref{fig:density} $(c)$, it can be seen that the empirical probability density of check nodes messages exchanged by the Self-Corrected Min-Sum
 decoding is very close to the Gaussian density of the (GD) condition. For the empirical probability density we taken into account only
 unerased ({\em i.e.} non zero)  check node messages. As discussed in the above section, the Self-Corrected Min-Sum decoding behaves
 as the Min-Sum decoding applied on unerased messages. It follows that the Self-Corrected Min-Sum decoding satisfies the three conditions
 (DG), (CNP) and (VNP) and therefore there should be no gap between its performance and the Sum-Product decoding performance. However,
 we should be careful about this assertion, because the recurrence relation of the error probability of the Self-Corrected Min-Sum decoding
 should be slightly different from the one of the Sum-Product decoding.
 In fact, if we see the Self-Corrected Min-Sum as the Min-Sum decoding applied on unerased messages,
 then the variable node degrees may vary from one iteration to another and this depends on the erasure probability. For instance, in Fig.~\ref{fig:tree},
 if $\alpha = \alpha_{m_1,n_1}$ is erased at some iteration, then the sub-tree ${\cal H}_1$ is discarded and the degree of
the
 variable node $n$ decreases by one.

 So far we have given a rather intuitive explanation of the observed behavior.
 In  order to write it formally, we need to split the error probability as
  $P_e^l = P^l + E^l$,
  where $P^l = \Pr(\alpha^l_{m,n} < 0)$ and $E^l = \Pr(\alpha^l_{m,n} = 0)$. The second probability is called the erasure probability at
  iteration $l$. We also note $R_e^l = \Pr(\beta^l_{m,n} \leq 0) = R^l + F^l$,
  where $R^l = \Pr(\beta^l_{m,n} < 0)$ and $R^l = \Pr(\beta^l_{m,n} = 0)$. The joint recurrence relation of $(P^l, E^l)$ is derived in
  several steps as follows.

  \smallskip\noindent $1)$ A check node message $\beta^{l+1}_{m,n}$ is erased if and only if there exists an erased incoming variable node
  messages $\alpha_{m,n'}^l = 0$. Since this happens with probability $(1-E^l)^{j-1}$, where $j$ is the degree of the check node $m$, and
  averaging over all possible check node degrees, we get:
  $$\begin{array}{r@{\ }c@{\ }l}
    F^{l+1} & = & \displaystyle\sum_{j=2}^{d_c}\rho_j\left(1-(1-E^l)^{j-1}\right)\\
            & = & \displaystyle1 - \rho(1-E^l)
    \end{array}$$

  \noindent $2)$ A check node message $\beta^{l+1}_{m,n} < 0$ if and only if all incoming variable node
  messages $\alpha_{m,n'}^l$ are not erased, and if the number of incoming negative messages is odd. Using \cite{gall-monograph}, lemma $4.1$, we get:
  $$\begin{array}{r@{\ }c@{\ }l}
   R^{l+1} &  = & \displaystyle\frac{1}{2} \sum_{j=2}^{d_c}\rho_j(1-E^l)^{j-1}\left(1 - \left(1-2\frac{P^l}{1-E^l}\right)^{j-1}\right)\\
           &  = & \displaystyle\frac{1}{2}\left(\rho(1-E^l) - \rho(1-E^l-2P^l)\right)
   \end{array}$$
   On the other hand, at iteration $l+1$, check node messages have a symmetric Gaussian distribution with mean $m_\beta^{l+1}$. It follows
   that:
   $$\begin{array}{r@{\ }c@{\ }l}
   R^{l+1} & = & \displaystyle Q\left(\sqrt{\frac{m_\beta^{l+1}}{2}}\right) \\
   m_\beta^{l+1} & = & 2Q^{-1}\left(R^{l+1}\right)^2
   \end{array}$$

\noindent $3)$ At iteration $l+1$, the temporary extrinsic LLR $\alpha_{m,n}^\tmp$ is the sum of the a priori information $\gamma_n$ and
the unerased incoming check node messages $\beta_{m,n}^{l+1} \not=0$. If the variable node $n$ is of degree $i$, the expected number of not
erased incoming check node messages is $(1 - F^{l+1})(i-1) = \rho(1-E^l)(i-1)$. Then the expectation of $\alpha_{m,n}^\tmp$ is $m_0 +
\rho(1-E^l)(i-1)m_\beta^{l+1}$ and it follows that:
 $$\Pr(\alpha_{m,n}^\tmp < 0) = \displaystyle Q\left(\sqrt{\frac{ m_0 + \rho(1-E^l)(i-1)m_\beta^{l+1} }{2}}\right) $$
 For convenience, we denote the above probability by $Q^{l+1}_i$.
 Furthermore, since the variable node message $\alpha_{m,n}^{l+1} < 0$ iff   $\alpha_{m,n}^{l} \leq
 0$ and $\alpha_{m,n}^\tmp < 0$, and
 averaging over all possible variable node degrees, we get:
 $$P^{l+1} = \displaystyle P^l_e\sum_{i=2}^{d_v}\lambda_i Q^{l+1}_i$$

 \noindent $4)$ With the above notation, the variable node message $\alpha_{m,n}^{l+1}$ is erased if and only if $\alpha_{m,n}^{l} <
 0$ and $\alpha_{m,n}^\tmp > 0$, or $\alpha_{m,n}^{l} > 0$ and $\alpha_{m,n}^\tmp < 0$. Consequently, we obtain:
 $$E^{l+1} = P^l(1-\sum\lambda_i Q^{l+1}_i) + (1 - P^l_e)\sum\lambda_i Q^{l+1}_i$$
 Moreover, we note that the following equality holds:
 $$P_e^{l+1} = P^l(1-\sum\lambda_i Q^{l+1}_i) + \sum\lambda_i Q^{l+1}_i$$

 For convenience, we introduce the following notation:
 $$\begin{array}{r@{\ }c@{\ }l}
 R(x,y)   & = & \displaystyle\frac{\rho(1-y)-\rho(1-y-2x)}{2} \\
 Q_i(x,y) & = & \topspace[8mm]\displaystyle Q\left(\sqrt{\frac{1}{\sigma_2} + \rho(1-y)(i-1)Q^{-1}\left(R(x,y)\right)}\right)
 \end{array}$$
 \begin{theo}
 Let $\phi = (\phi_1, \phi_2)$, where:
  $$\begin{array}{r@{\ }c@{\ }l}
  \phi_1(x,y) & = & \displaystyle(x+y)\sum_{i=2}^{d_v}\lambda_i Q_i(x,y) \\
  \phi_2(x,y) & = & \displaystyle x + (1 - y - 2x)\sum_{i=2}^{d_v}\lambda_i Q_i(x,y)
  \end{array}$$
  Then the following recurrence relation holds:
  $$(P^{l+1}, E^{l+1}) = \phi(P^l, E^l)$$
 \end{theo}

 \section{Simulation results} \label{sec:simulations}
 We present Monte-Carlo simulation results for coding rate $1/2$ over the AWGN channel with QPSK modulation.

\begin{figure}[h]
 \includegraphics[width=\linewidth]{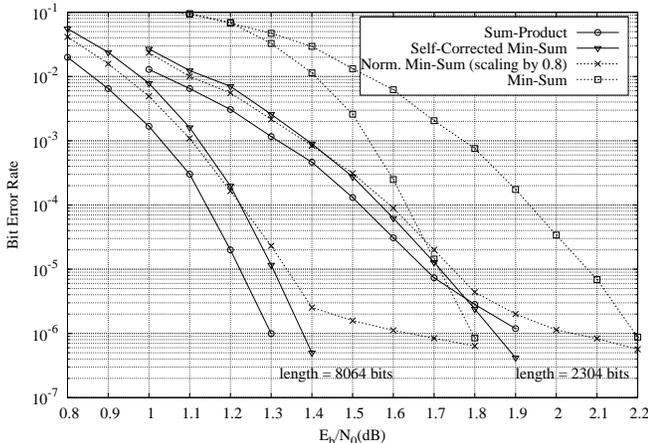}

 \vspace{-5.5mm}
 \caption{Irregular LDPC codes, floating point simulation, Bit Error Rate, max iter number  = 200}
 \label{fig:scmc_courbeBER}
 \end{figure}

\begin{figure}[t]
 \includegraphics[width=\linewidth]{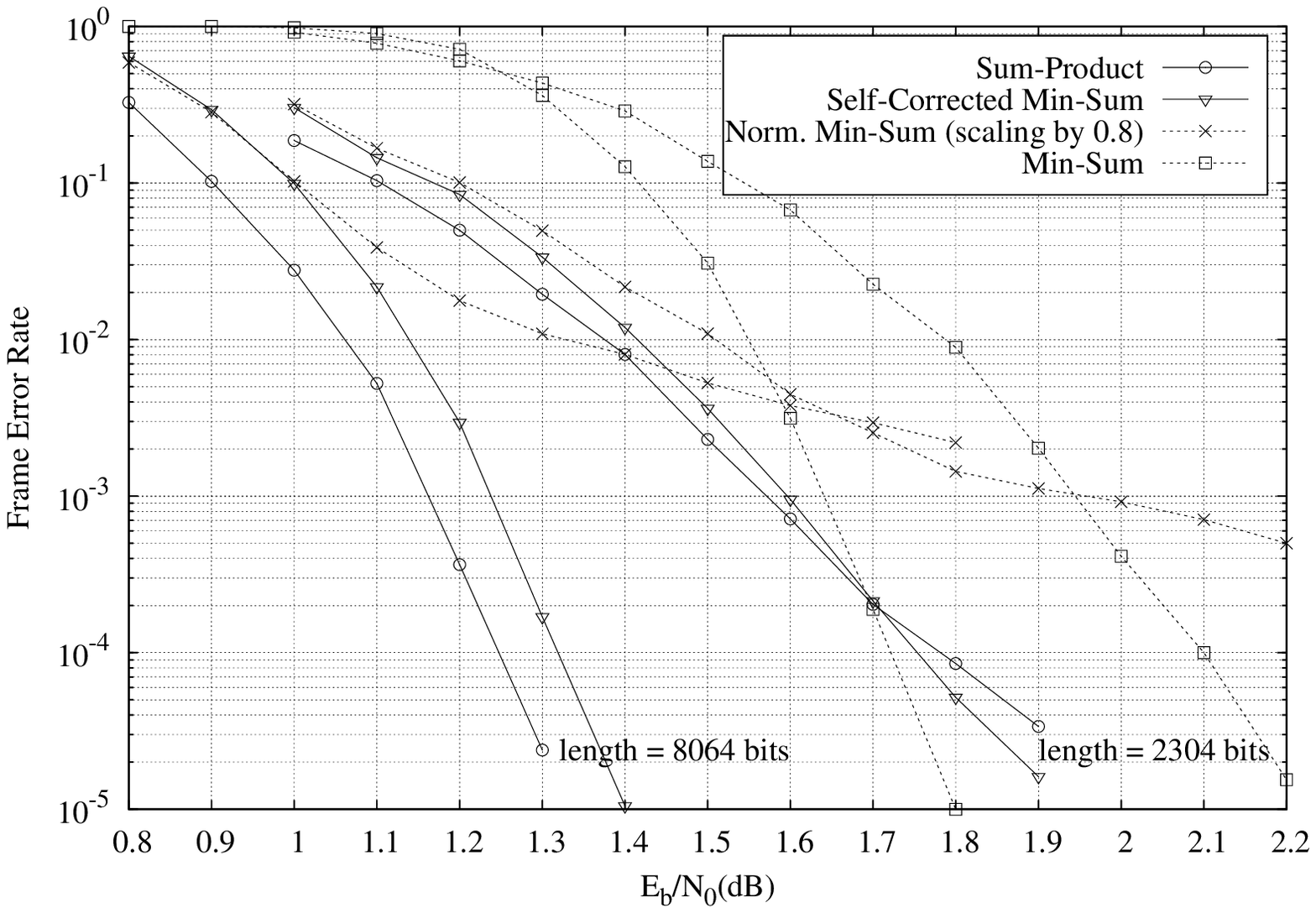}

 \vspace{-5.5mm}
 \caption{Irregular LDPC codes, floating point simulation, Frame Error Rate, max iter number  = 200}
 \label{fig:scmc_courbeFER}
 \end{figure}

Floating-point performance of optimized irregular LDPC codes under the Sum-Product, Min-Sum, Self-Corrected Min-Sum, and  Normalized
Min-Sum decoders is presented in Fig.~\ref{fig:scmc_courbeBER} and Fig.~\ref{fig:scmc_courbeFER} in terms of bit and frame error rates,
respectively. We note that the performance gap between the Self-Corrected Min-Sum  and the Sum-Product decoders is only of  $0.05$~dB for
short code lengths ($2304$ bits) and of $0.1$~dB for long code lengths ($8064$ bits). Moreover, in case of short code lengths the
Self-Corrected Min-Sum decoder outperforms the Sum-Product decoder at high SNR values, corresponding to the error floor of the second
decoder. Concerning the Normalized Min-Sum decoder, its bit error rate performance is very close to the one of the Self-Corrected Min-Sum
decoder at low SNR values; however, it presents an error floor at higher SNR values. Furthermore, it behaves completely faulty in terms of
frame error rate, which is quite surprising as this does not fit with the bit error rate behavior.

\begin{figure}[t]
 \vspace{-1mm}
 \includegraphics[width=\linewidth]{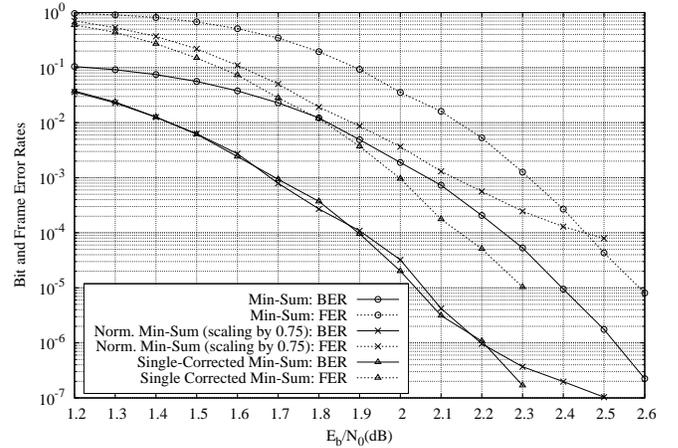}

 \vspace{-5.5mm}
 \caption{IEEE-$802.16$e LDPC codes, fixed point simulation ($-8\leq \gamma,\alpha,\beta< 8$ and $-32\leq\tilde\gamma< 32$, with quantization step $0.25$), max iter number = 30}\label{fig:wimax_courbe}
 \end{figure}

Fig \ref{fig:wimax_courbe} presents fixed point simulation results for the quasi-cyclic LDPC codes of length $2304$ bits from the
IEEE-$802.16$e standard \cite{wimax_std}. The above conclusions also apply in this case.

  \vspace{-3.5mm}

 \section{Conclusions} \label{sec:conclusions}

A very simple but powerful self-correction method for the Min-Sum decoding of LDPC codes was presented in this paper. The  proposed
Self-Corrected Min-Sum algorithm performs quasi-optimal iterative decoding, while preserving low Min-Sum complexity and independence with
respect to noise variance estimation errors. This makes the Self-Corrected Min-Sum decoding very attractive for practical purposes.

 \bibliographystyle{abbrv}
\vspace{-1mm}
\bibliography{./bib/MyBiblio,./bib/Zotero}

\end{document}